# Emergence of metallic monoclinic states of VO$_2$ films induced by K deposition

D. Shiga,[1,2] M. Minohara,[2] M. Kitamura,[2] R. Yukawa,[2] K. Horiba,[2,*] and H. Kumigashira[1,2]

[1]*Institute of Multidisciplinary Research for Advanced Materials (IMRAM), Tohoku University, Sendai 980-8577, Japan*

[2]*Photon Factory, Institute of Materials Structure Science, High Energy Accelerator Research Organization (KEK), 1-1 Oho, Tsukuba 305-0801, Japan*

**Abstract**

In order to study the origin of metallization of VO$_2$ induced by electron injection, we deposit K atoms onto the surface of VO$_2$ films grown on TiO$_2$ (001) substrates, and we investigate the change in the electronic and crystal structures using *in situ* photoemission spectroscopy and x-ray absorption spectroscopy (XAS). The deposition of K atoms onto a surface of insulating monoclinic VO$_2$ leads to a phase transition from insulator to metal. In this metallization state, the V-V dimerization characteristic to the monoclinic phase of VO$_2$ still exists, as revealed by the polarization dependence of the XAS spectra. Furthermore, the monoclinic metal undergoes a transition to a monoclinic insulator with decrease in temperature, and to a rutile metal with increase in temperature. These results indicate the existence of a metallic monoclinic phase around the boundary between the insulating monoclinic and metallic rutile phases in the case of electron-doped VO$_2$.

*Corresponding author: horiba@post.kek.jp

# I. INTRODUCTION

Metal–insulator transitions (MITs) are one of the most fascinating physical phenomena to be observed in strongly correlated oxides [1]. MIT is accompanied by an orders-of-magnitude change in conductivity that is induced by external stimuli, and this phenomenon has formed a central topic in modern condensed matter physics for its potential application in future electronic devices [2,3]; further, the phenomenon provides an opportunity to better understand the fundamental physics of strongly correlated oxides. Among the MITs exhibited by strongly correlated oxides, the MIT of vanadium dioxide ($VO_2$) [4] is particularly intriguing because both the structural transition and electron correlation contribute to the MIT: The structural phase changes from metallic rutile ($P4_2/mnm$) to insulating monoclinic ($P2_1/c$) across the MIT at nearly room temperature (see Fig. 1) [5]. As shown in the inset of Fig. 1, tilting and pairing of V ions along the $c_R$-axis, which is defined as the $c$-axis of the rutile structure, in the monoclinic phase mark this structural change. Since the distances between paired ions and between ion pairs are different, the V ions in $VO_2$ are dimerized along the $c_R$-axis in the insulating monoclinic phase [5,6]. Although the MIT that is concomitant with the dimerization of V atoms is reminiscent of the Peierls transition [7,8], the importance of strong electron correlations in $VO_2$ has also been evident for this MIT as per a large number of experimental and theoretical investigations [9–11]. Therefore, the mechanism of the MIT in $VO_2$ is now mainly understood as a collaborative Mott–Peierls (or Peierls–Mott) transition [12–16].

This type of MIT driven by the cooperation of the two instabilities has motivated researchers to control the phenomenon via external stimuli such as light [17], strain [18,19], or electrostatic injection of charge (electron doping) using field-effect transistor (FET) structures [20–23]. In this regard, Nakano *et al*. have demonstrated that the low-temperature insulating state of $VO_2$ can be completely switched over to the first-order MIT by application of a gate voltage [20]. Subsequently, Jeong *et al*. have also reported on gate-induced metallization in $VO_2$ via the same FET technique using ionic liquid; however, they have claimed that the metallization originates from electric-field induced oxygen vacancy formation and not electrostatic surface charge accumulation [21]. Recently, Li *et al*. have reported the stabilization of metal-like monoclinic intermediate states via the surface coordination effect of electron-donor molecules on the VO$_2$ surface [24]. The possible formation of the third phase, the metallic monoclinic phase, around the boundary between the metallic rutile and insulating monoclinic phases in VO$_2$ under the application of external stimuli has also been suggested by several researchers [24–28]. Hence, the underlying physical mechanism of the electric-field-induced metallization process is still under debate despite subsequent intense experimental efforts [20–23]. The key issues to be clarified are whether the electrostatic carrier injection (electron transfer) results in the suppression of the MIT temperature $T_{\mathrm{MI}}$ and whether the initially insulating VO$_2$ remains in the monoclinic phase even after metallization.

To understand the origin of the MIT induced by electron doping and the possible formation of the intermediate metallic monoclinic phase, it is crucial to obtain information on how the electronic structure and characteristic dimerization of VO$_2$ change through the metallization phenomena. Against this backdrop, in this study, we performed electron doping of a VO$_2$ surface via *in situ* deposition of K atoms, wherein the K atoms were chemically adsorbed at the surface and the adsorbed K atoms doped electrons into the surface region [29–32]. Subsequently, we studied the resulting structural and electronic changes via *in situ* photoemission spectroscopy (PES) and x-ray absorption spectroscopy (XAS) measurements. K-atom deposition onto a surface of insulating monoclinic VO$_2$ caused a phase transition from insulator to metal, which was evident by the appearance of a clear Fermi-edge cutoff in the PES spectra, thereby indicating the occurrence of the MIT via surface carrier accumulation. In this metallization state, the features of the dimerization characteristic to the monoclinic phase of VO$_2$ were revealed via the polarization dependence of the XAS spectra. Furthermore, the fingerprint of the monoclinic phase disappeared with increase in temperature, although the metallic state itself continued to exist. These results indicate the existence of a metallic monoclinic phase around the boundary between the insulating monoclinic and metallic rutile phases in the case of electron-doped VO$_2$.

## II. EXPERIMENT

As regards sample preparation, oxide-film growth was first performed in a pulsed laser deposition (PLD) chamber, and K deposition was subsequently carried out in a preparation chamber. These chambers were connected to a synchrotron-radiation PES system at the undulator beamline BL-2A MUSASHI of the Photon Factory, KEK [33]. $VO_2$ films with a thickness of approximately 12 nm were epitaxially grown on 0.05 wt. % Nb-doped rutile $TiO_2$ (001) substrates via PLD [18,34,35]. During $VO_2$ deposition, the substrate temperature was maintained at 400°C and the oxygen pressure was maintained at 10 mTorr [35]. Here, we remark that we carefully optimized the growth temperature to avoid interdiffusion of the constituent transition metals across the interface by measuring the temperature-dependent resistivity ($\rho$–$T$) of the grown $VO_2$ films [18]. The surface structure and cleanness of the vacuum-transferred $VO_2$ films were confirmed via low-energy electron diffraction (LEED) and core-level photoemission measurements. The LEED pattern showed sharp 1 × 1 spots, thereby confirming the high surface crystallinity and cleanness essential for our spectroscopic measurements. No detectable C 1$s$ signal was observed in the core-level photoemission spectra. The results of the detailed characterization of the grown $VO_2$ films are presented in the Supplemental Material [36].

After the oxide films were grown, K atoms were deposited onto them via a K dispenser. The K-deposition rate was monitored by means of a quartz microbalance as well as by analyzing the relative intensities of the relevant core levels [32]. The surface morphologies of the measured films were analyzed via *ex situ* atomic force microscopy in air. The crystal structures of the $VO_2$ films were characterized via x-ray diffraction, which confirmed the epitaxial growth of the oxide films on the substrates. The electrical resistivity was measured by means of the standard four-probe method. $T_{MI}$ was determined to be 295 K, while the change in the resistivity across the MIT [$\rho$ (250 K)/ $\rho$ (320 K)] was considerably larger than $10^3$ [36]. These values are almost the same as the corresponding ones of the reported epitaxial $VO_2$ films grown on $TiO_2$ (001) substrates under in-plane tensile strain [18].

PES measurements were performed *in situ* with the use of a VG-Scienta SES-2002 analyzer with a total energy resolution of 120 meV and 200 meV at photon energies of 700 eV and 1200 eV, respectively. The XAS spectra were also measured *in situ* with linearly polarized light via measurement of the sample drain current. For linear dichroism (LD) measurements, we acquired the XAS spectra at angles $\theta = 0°$ and $60°$ between the $a_R$-axis direction and the polarization vector while maintaining a fixed angle between the direction normal to the surface and the incident light (see Supplemental Material [36]). The Fermi level ($E_F$) of each sample was determined by measurement of a gold film that was electrically connected to it. Because it is known that $VO_2$ exhibits MIT upon irradiation by light [17], we paid special attention to the possible spectral change induced by the light irradiation. Owing to the relatively wide spot size (200 μm × 500 μm) of the synchrotron light at the Photon Factory, KEK, there were no detectable spectral differences before and after 30 min of synchrotron-light irradiation. Since certain slight changes were observed in the spectra after 2 h of irradiation by soft-x-ray light, we changed the location of the light spot on each sample every 30 min during the spectroscopic measurements. Furthermore, to prevent the oxidation of the deposited K atoms and to avoid the resulting appearance of spurious peaks in the spectra, the oxide-film growth, K deposition, and subsequent spectroscopic measurements were performed *in situ*. All the PES and XAS spectra were acquired at the temperatures indicated by circles in Fig. 1.

## III. RESULTS AND DISCUSSION

### A. PES results

Figure 2 shows the temperature dependence of the valence-band spectra of the VO$_2$ films before and after K deposition. These spectra mainly consist of two features: structures derived from O 2$p$ states at binding energies of 3–10 eV and peaks derived from the V 3$d$ states near $E_F$ [37–39]. As regards the bare VO$_2$ films before K deposition, as shown in Fig. 2(a), the spectra exhibit the characteristic features representative of the MIT of VO$_2$: The spectrum near $E_F$ in the metallic phase (measured at $T$ = 320 K) consists of a sharp coherent peak just at $E_F$ and a weak broad satellite structure around 1.2 eV, while that in the insulating phase (at $T$ = 250 K) shows a single peak around 1.1 eV, which leads to the formation of an energy gap at $E_F$. Furthermore, focusing on the O 2$p$ states, we observe dramatic changes across the MIT. These changes are responsible for the structural change concomitant with the MIT in VO$_2$ [11]. Although the $T_{MI}$ of the VO$_2$ film is reduced to ~295 K owing to the in-plane tensile strain effect [18], the spectral change across the MIT is in excellent agreement with the results of previous studies on single-crystal VO$_2$ [11], which indicates that the VO$_2$ films in the present work are of nearly the same quality as single-crystal samples.

When K atoms are deposited on the surface of VO$_2$ films in the insulating phase at 250 K, the V 3$d$ states near $E_F$ exhibit the dramatic changes. The single sharp peak around 1.1 eV in insulating VO$_2$ films [see Fig. 2(a)] shifts to a lower binding energy of 0.6 eV and becomes significantly broader in the spectra of VO$_2$ films after K deposition (K/VO$_2$) at 250 K, as shown in Fig. 2(b). Meanwhile, a closer examination of the spectral shape near $E_F$ reveals the appearance of a distinct Fermi-edge profile in K/VO$_2$, thus indicating the metallization of VO$_2$ films by K deposition on the surface. K atoms are expected to be chemically adsorbed onto the VO$_2$ surfaces and the adsorbed atoms dope electrons into the surface region, as in the case of K adsorption onto TiO$_2$ surfaces [32] as well as on other oxide surfaces [29–31]. Therefore, the appearance of the Fermi-edge profile after K deposition on the insulating VO$_2$ films suggests that the transition from insulator to metal is induced by electron doping.

The appearance of the Fermi-edge profile in the K/VO$_2$ spectra concomitant with predominant satellite structures suggests the emergence of unusual metallic states in K/VO$_2$ films: The coherent states at $E_F$ are significantly suppressed in comparison with that in the metallic phase of bare VO$_2$ [see the spectra at 320 K in Fig. 2(a)]. To investigate the metallic states in more detail, we measured the temperature dependence of K/VO$_2$ films, as shown in Fig. 2(b). With decrease in temperature to $T$ = 150 K, the satellite structure at 0.6 eV shifts to a higher binding energy of 1.1 eV, resulting in the opening of an energy gap at $E_F$, thereby indicating the occurrence of temperature-induced MIT in K/VO$_2$ films at $T_{MI}$ values of 150–250 K. Although the peak of the insulating K/VO$_2$ films is relatively broader than that of insulating bare VO$_2$ films, the nearly unchanged peak position of the lower Hubbard band suggests that the ground states of K/VO$_2$ are the same as those of VO$_2$ films. In other words, the $T_{MI}$ value of VO$_2$ films is suppressed to within 150–250 K by K deposition. Since the K atoms adsorbed onto the VO$_2$ surface dope electrons into the surface region [29–32], these results suggest that the electron doping into the surface region leads to the MIT of VO$_2$, as in the case of electrostatic surface charge accumulation in electric-double-layer transistors (EDLTs) based on VO$_2$ [20]. It should be noted that the influence of H-induced metallization of VO$_2$ in the EDLT process [23] should be ignored in the present study, since K atoms are deposited *in situ* on the vacuum-transferred VO$_2$ films under ultrahigh vacuum conditions.

Here, it is worth evaluating the carrier number induced by K deposition. In this regard, Shibuya *et al.* have reported an unusual phase diagram of epitaxially stabilized V$_{1–x}$W$_x$O$_2$ films, wherein electron doping is caused by the substitution of W$^{6+}$ ions for V$^{4+}$ ions [35]. In these electron-doped VO$_2$ films, $T_{MI}$ monotonically decreases with increasing $x$ (corresponding to $2x$ electrons per unit cell), and eventually, a metallic ground state is stabilized at $0.08 \leq x \leq 0.09$ (electron doping of 0.16–0.18 per unit cell), as indicated in Fig. 1 by the dotted line. Under the assumption that the effect of electron doping by K deposition is equivalent to that by the chemical substitution, the modulation of $T_{MI}$ in K/VO$_2$ may correspond to an electron doping level of 0.04–0.10 per V atom. This value is of the same order as the maximum amount of electron doping induced by K adsorption onto other strongly correlated oxides [30]. These results suggest that the carrier-induced MIT is realized by surface carrier injection from K into $VO_2$ films.

On the other hand, when the measurement temperature is raised to 320 K from 250 K, the coherent states at $E_F$ evolve into a sharp peak structure reminiscent of that in the metallic rutile phase of $VO_2$. The V 3$d$ states near $E_F$ consist of a sharp coherent peak at $E_F$ and a broad satellite structure around 1.2 eV, although the relative intensity of the coherent peak is weaker in K/$VO_2$ films. This means that the metallization states of K/$VO_2$ at 250 K might be different from the metallic phase of bare $VO_2$, as well as the high-temperature phase of K/$VO_2$. These results of the temperature dependence of the spectral shape for K/$VO_2$ suggest that a certain different metallic phase exists at the phase boundary near the MIT in the electron-doped K/$VO_2$. To further investigate these phase transition phenomena from the viewpoint of crystal structures, we next discuss the possible formation of a *metallic monoclinic phase* at the boundary of the MIT in connection with the polarization-dependent XAS results.

### B. Polarization-dependent XAS results

On the basis of the PES results, we call attention to the possibility that some other metallic phase is formed at the boundary of the MIT of electron-doped $VO_2$. Thus, the next crucial issue is to determine the relationship between the carrier-induced metallization at 250 K and the dimerization of V ions. The structural change characteristic of the MIT in $VO_2$ has previously been revealed by the polarization dependence of oxygen $K$-edge XAS (O $K$ XAS) [8,11,40–42]. The XAS at the O $K$ edge is a technique complementary to PES for investigating the conduction band via probing of the unoccupied O 2$p$ partial density of states that are mixed with the unoccupied V 3$d$ states. Because dimerization in the monoclinic phase splits a half-filled $d_{//}$ state into occupied $d_{//}$ and unoccupied $d_{//}^*$ states [6], an additional peak corresponding to the $d_{//}^*$ states appears in the XAS spectra only for the insulating monoclinic phase of $VO_2$, as can be observed in Fig. 3(a). Furthermore, the observed additional peak can be identified with the $d_{//}^*$ state by inferring the polarization dependence (i.e., linear dichroism) of the XAS spectra [11,42]: The matrix elements of the dipole transition for the $d_{//}^*$ state completely disappear for the geometry of $E \perp c_R$ axis (i.e., the direction along the one-dimensional V-V atomic chains) owing to the orbital symmetry of $d_{//}^*$. In fact, as can be confirmed from Fig. 3(a), the $d_{//}^*$ states are observed at 530.7 eV (indicated by the solid triangle) in insulating monoclinic $VO_2$ films at $T = 250$ K, whereas they disappear in metallic rutile ones at $T = 320$ K. Thus, the appearance of $d_{//}^*$ in the LD spectra can be used as a fingerprint of the monoclinic structure with the V-V dimer in $VO_2$.

In order to investigate the role of dimerization in the MIT of K/$VO_2$ films, we carried out measurements of the polarization dependence of O $K$ XAS. Figure 3(b) shows the polarization dependence of the O $K$ XAS spectra for K/$VO_2$ at each measurement temperature and their LD spectra. An additional shoulder structure is observed around 530.6 eV for K/$VO_2$ films measured at $T = 250$ K and 150 K. Notably, the additional structure is observed only in the $E // c_R$ spectra and not in the $E \perp c_R$ spectra. The absence of the additional peak in the $E \perp c_R$ spectra indicates that the peak originates from the $d_{//}^*$ state due to V-V dimerization. Owing to its symmetry, the $d_{//}^*$ state lacks hybridization with O $2p$ orbitals perpendicular to the $c_R$ axis. The emergence of the $d_{//}^*$ state for K/$VO_2$ films measured below $T = 250$ K is highlighted in their LD spectra. These results indicate that the structural phase transition in K/$VO_2$ occurs in the temperature range of 250–320 K.

Upon comparison of the O $K$ XAS spectra of K/$VO_2$ with those of bare $VO_2$ films, the overall spectral features of K/$VO_2$ at $T = 250$ K and 150 K exhibit a close similarity with those of the insulating monoclinic phase of bare $VO_2$, while the K/$VO_2$ spectrum at 320 K is in excellent agreement with that of the metallic rutile phase of bare $VO_2$. In connection with the PES results shown in Fig. 2, it is naturally concluded that K/$VO_2$ at 320 K and 150 K is in the metallic rutile and insulating monoclinic phases, respectively. Meanwhile, K/$VO_2$ at 250 K is assigned as the monoclinic phase from the LD-XAS results, whereas it shows metallic behavior in the PES spectra. That is, considering both the PES and XAS results, it is suggested that K/$VO_2$ at $T = 250$ K can be assigned as a *"monoclinic metal"* having the metallic electronic state [see Fig. 2 (b)] and a dimerized monoclinic crystal structure [see Fig. 3(b)]. These spectroscopic results also suggest that an additional metallic monoclinic phase exists at the boundary between the metallic rutile phase at high temperature and the insulating monoclinic phase at low temperature in the electron-doped $VO_2$.

### C. Verification of metallic monoclinic phase of electron-doped $VO_2$

Although the spectroscopic results suggest the formation of a carrier-induced "*metallic monoclinic phase*" in K/$VO_2$ at the boundary of the MIT, there are two important questions to be clarified. One is the possibility of phase separation in K/$VO_2$ and the other is the possible influence of oxygen-vacancy formation in $VO_2$ associated with K deposition. For the former, it is well known that $VO_2$ films under external stimuli exhibit a spatial phase separation wherein the metallic rutile domains in the micrometer scale coexist with the insulating monoclinic ones [14,16,43,44]. Since the PES and XAS measurements capture the average information of the two phase-separated phases owing to the significantly wider spot size (200 μm × 500 μm) of the synchrotron light beam, there is the possibility that the metallic states originate from the results of the existence of the two domains. In order to verify the possible phase separation, we attempted to reproduce the PES spectrum near $E_F$ of the "monoclinic metal" phase of K/$VO_2$ (250 K) via a linear combination of the spectra for the metallic rutile phase (320 K) and insulating monoclinic phase (150 K) of K/$VO_2$. The best-fit results are shown in Fig. 4. As can be seen in the Fig. 4, it appears to be difficult to reproduce the spectrum by means of a simple linear combination. From these results, it is concluded that the "*metallic monoclinic phase*" is formed at the boundary of the MIT between the insulating monoclinic phase and metallic rutile phase in K/$VO_2$ films.

Finally, we discuss the other possibility that the metallization states in K/$VO_2$ are caused by oxygen deficiency due to K deposition. There are several studies that have claimed that metallization in EDLTs originates from electric-field-induced oxygen vacancy formation and not electrostatic surface charge accumulation [21,45,46]: The metallization states remain unchanged after washing out of the ion liquid (removing the electrostatic gating) [21,47]. These results drive us to the question as to whether the deposition of K atoms onto the $VO_2$ surface leads to oxygen vacancy owing to the high chemical activity of K. To verify the possibility of oxygen-vacancy-induced metallization in K/VO$_2$, we oxidized the absorbed K atoms by exposing the sample to air [48]; the corresponding results are shown in Fig. 5. From the figure, we note that the metallic monoclinic K/VO$_2$ undergoes phase transition to insulator upon oxidation of adsorbed K atoms. Furthermore, the spectra of K/VO$_2$ after air exposure exhibit the features characteristic to the insulating monoclinic phase of VO$_2$: The dominant peak locates at a higher binding energy of 1.2 eV, resulting in an energy gap opening up at $E_\text{F}$. Since electrons doped from K atoms adsorbed onto VO$_2$ are removed by the oxidation of the K atoms, these results indicate that the observed "*metallic monoclinic phase*" in K/VO$_2$ is induced by carrier accumulation in VO$_2$.

## IV. CONCLUSION

We performed *in situ* PES and XAS measurements to investigate the change in the electronic and crystal structures of $VO_2$ epitaxial films induced by carrier injection. For surface carrier injection into $VO_2$, K atoms were deposited *in situ* onto $VO_2$ film surfaces, wherein the K atoms were chemically adsorbed at the surface and the adsorbed K atoms doped electrons into the surface region. By the oxidation process of the adsorbed K atoms, it was confirmed that the adsorbed K atoms act as electron donor to the surface of $VO_2$. The deposition of K atoms onto the surface of insulating monoclinic $VO_2$ led to a phase transition from insulator to metal, thereby suggesting that the $VO_2$ film in the insulating phase was successfully metallized by surface carrier injection upon K deposition. Notably, in this metallization state, the V-V dimerization characteristic to the monoclinic phase of $VO_2$ still existed, thus indicating the existence of a new metallic phase maintaining the V-V dimer structure, termed the *metallic monoclinic phase*. Furthermore, the monoclinic metal underwent a transition to monoclinic insulator with decrease in temperature, and to rutile metal with increase in temperature. These results indicate the existence of a metallic monoclinic phase around the boundary between the insulating monoclinic and metallic rutile phases in electron-doped $VO_2$.

## ACKNOWLEDGMENTS

The authors are very grateful to A. F. Santander-Syro, M. Lippmaa, and M. J. Rozenberg for the useful discussions. This work was financially supported by Grants-in-Aid for Scientific Research (Nos. 16H02115, 16KK0107, 17K14325, and 16K05033) from the Japan Society for the Promotion of Science (JSPS), CREST (JPMJCR18T1) from the Japan Science and Technology Agency (JST), and the MEXT Elements Strategy Initiative to Form Core Research Center. This work at KEK-PF was performed under the approval of the Program Advisory Committee (Proposals 2015S2-005, 2016G164, and 2018S2-004) at the Institute of Materials Structure Science at KEK.

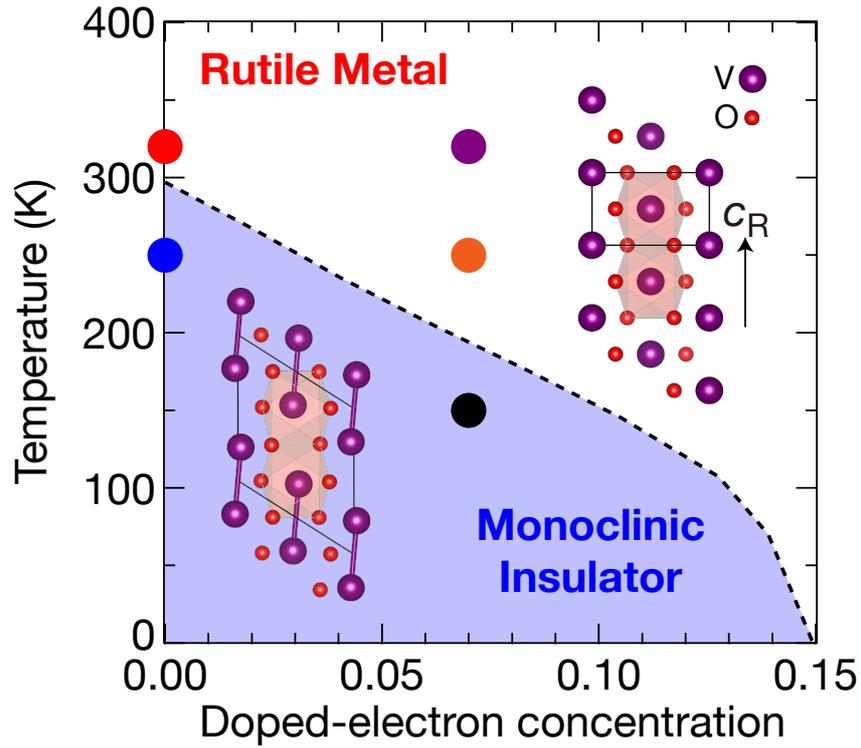

FIG. 1. (Color online) Possible electronic phase diagram of electron-doped $VO_2$ (001) films represented as functions of temperature and amount of electron doping. The dotted line is obtained from the results of the electronic phase diagram of epitaxially stabilized $V_{1-x}W_xO_2$ films [35] under the assumption that the effect of electron doping induced by K deposition is equivalent to that by chemical substitution of $W^{6+}$ for $V^{4+}$ ions. Colored solid circles represent spectroscopic measurement points. The inset shows the crystal structure of rutile (right top side) and monoclinic (left bottom side) $VO_2$.

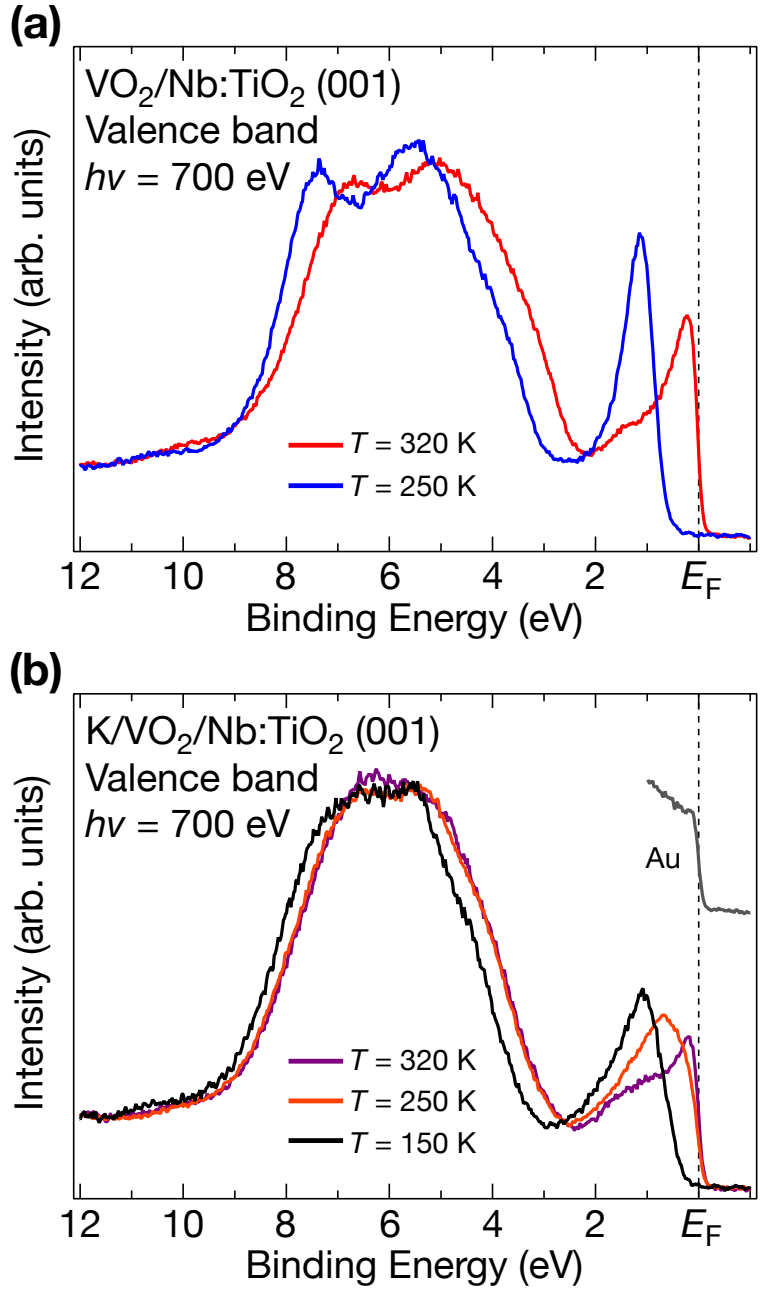

FIG. 2. (Color online) (a) Temperature dependence of valence-band PES spectra measured at $h\nu = 700$ eV for $VO_2$/Nb:$TiO_2$ (001) films before K deposition. (b) That after K deposition, wherein the spectrum near $E_F$ of Au acquired at $T = 250$ K under the same experimental conditions is shown as a reference of the Fermi-edge cutoff. Note that the colors of each spectrum correspond to those of the solid circles in Fig. 1.

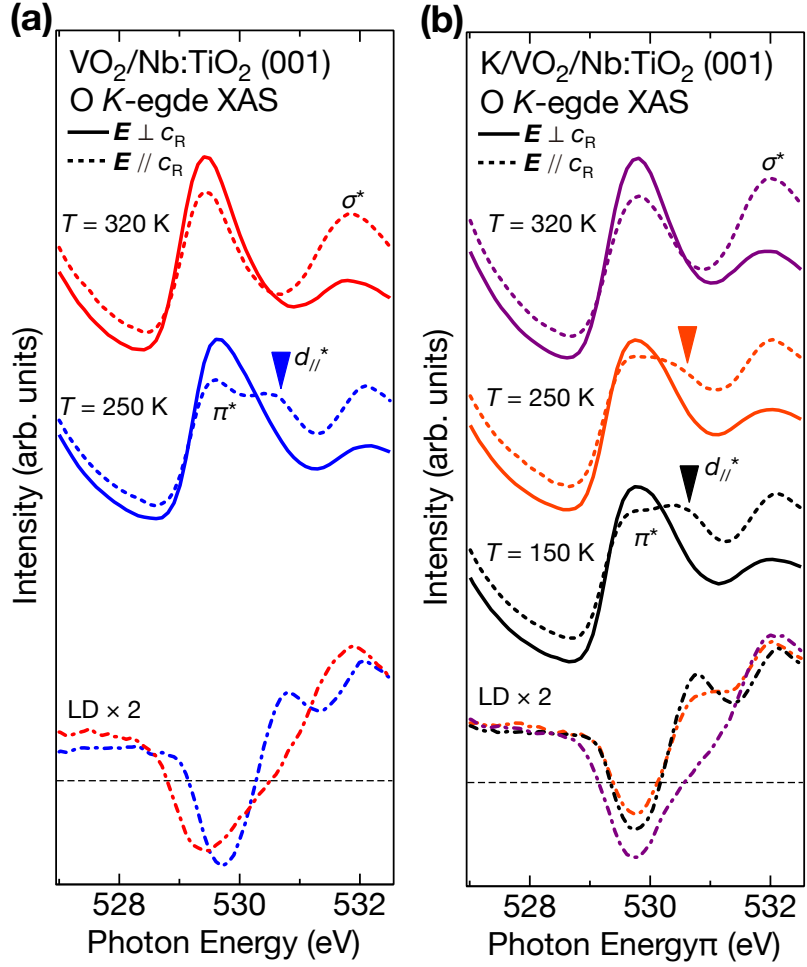

FIG. 3. (Color online) Temperature dependence of O $K$ XAS spectra with different polarizations (upper panel) and their LD spectra (lower panel) of $VO_2$/Nb:$TiO_2$ (001) films (a) before and (b) after K deposition. XAS spectra acquired with the polarization vector $E$ perpendicular to the $c_R$-axis ($E \perp c_R$) and parallel to the $c_R$-axis ($E // c_R$) are represented by solid and broken lines, respectively. XAS spectra with $E // c_R$ ($I_{//}$) are deduced from the expression $I_{//} = (4/3)(I - I_\perp/4)$, where $I_\perp$ (namely, that corresponding to $E \perp c_R$) and $I$ denote XAS spectra measured with normal ($\theta = 0°$) and grazing ($\theta = 60°$) incidence, respectively. Filled triangles indicate the position of shoulder structures originating from the $d_{//}^*$ states. Note that the colors of each spectrum correspond to those of the solid circles in Fig. 1. Following the assignments made in previous studies [11], the first peak around 529.5 eV can be assigned to $\pi^*$ bands formed by V $3d_{xz}$ and $3d_{yz}$ orbitals, while the second peak around 532 eV can be assigned to $\sigma^*$ bands formed by $3d_{z2-r2}$ and $3d_{x2-y2}$ orbitals.

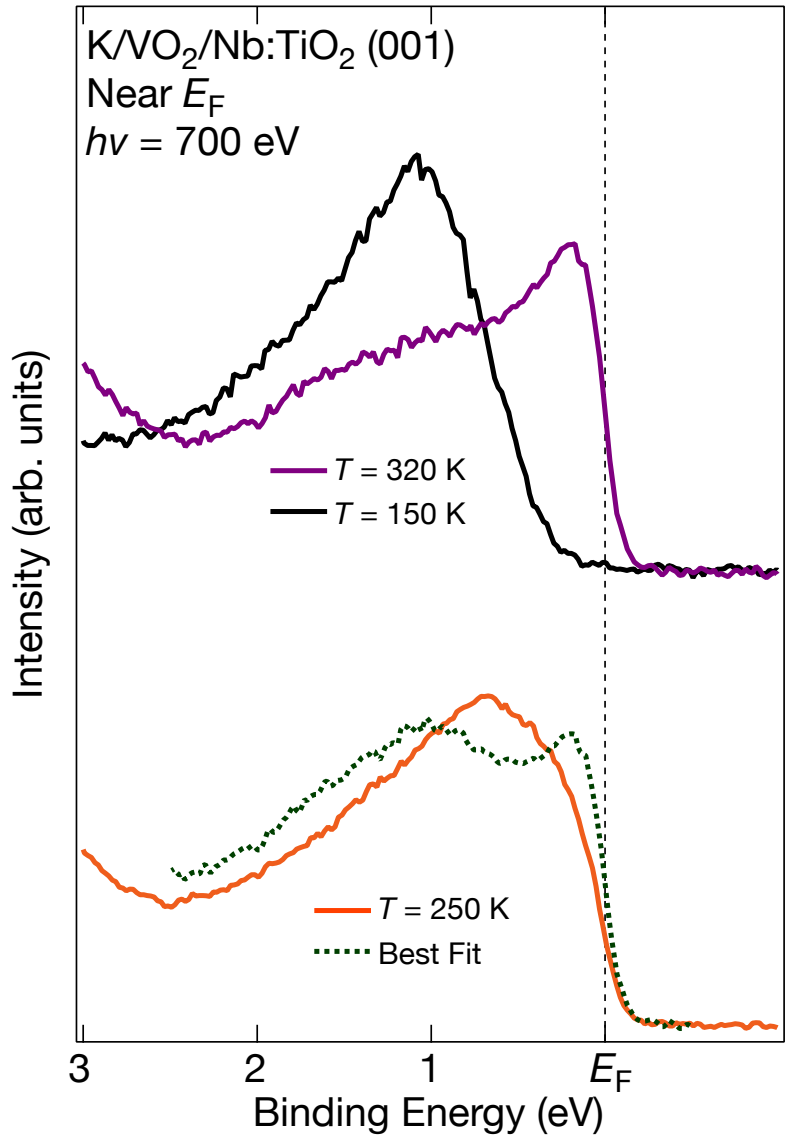

FIG. 4. (Color online) PES spectra near $E_F$ of metallic rutile (320 K) and insulating monoclinic (150 K) K/VO$_2$ films (upper panel) and comparison of "metallic monoclinic" K/VO$_2$ spectrum (250 K) with a linear combination of insulating monoclinic and metallic rutile K/VO$_2$ spectra (lower panel).

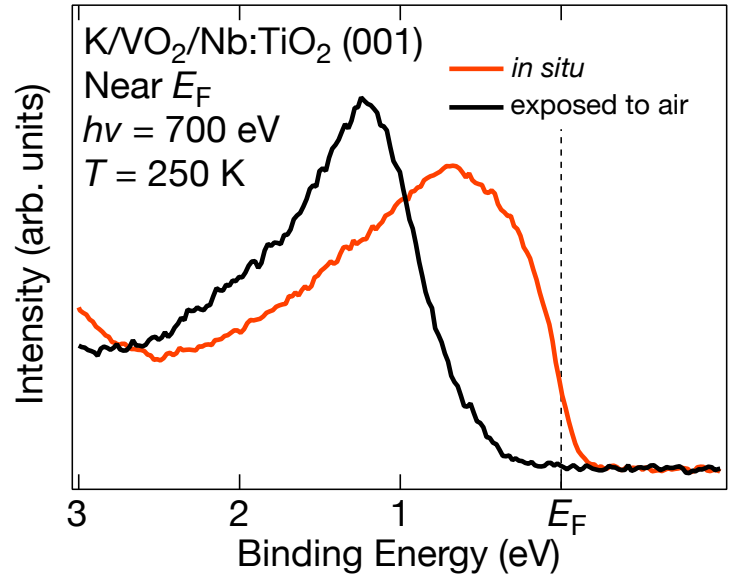

FIG. 5. (Color online) PES spectra near $E_F$ of K/VO$_2$ films before and after air exposure.